\documentclass[prl,twocolumn,epsfig]{revtex4}
\usepackage{graphics}
\usepackage{epsfig}
\usepackage{amsfonts}
\usepackage{amsmath}
\usepackage{bm}

\usepackage{color}
\begin{document}
\title{
Individual low-energy toroidal dipole state in $^{24}$Mg}
\author{V.O. Nesterenko$^1$,  A. Repko $^{2}$, J. Kvasil$^3$, and P.-G. Reinhard$^4$}
\affiliation{$^1$ Laboratory of Theoretical Physics, Joint Institute for
Nuclear Research, Dubna, Moscow region, 141980, Russia}
\affiliation{$^2$
Department of Nuclear Physics, Institute of Physics SAS, 84511, Bratislava, Slovakia}
\affiliation{$^3$
Institute of Particle and Nuclear Physics, Charles University, CZ-18000, Prague, Czech Republic}
\affiliation{$^4$
Institut f\"ur Theoretische Physik II, Universit\"at Erlangen, D-91058, Erlangen, Germany}
\date{\today}

\begin{abstract}
The low-energy dipole excitations in $^{24}$Mg are investigated within the Skyrme
quasiparticle random-phase-approximation (QRPA) for axial nuclei. The
calculations with the force SLy6 reveal a remarkable feature: the lowest $I^{\pi}K=1^-1$
excitation (E = 7.92 MeV) in $^{24}$Mg is a vortical toroidal state (TS)
representing a specific vortex-antivortex  realization of well-known spherical
Hill's vortex in a strongly deformed axial confinement. This is a striking
example of an {\it individual} TS which
can be much easier discriminated in experiment than the toroidal
dipole resonance embracing many states. The TS acquires the lowest energy due to
the huge prolate axial deformation in $^{24}$Mg. The result persists for different
Skyrme parameterizations (SLy6, SVbas, SkM*). We analyze spectroscopic
properties of the TS and its relation with the cluster structure of
$^{24}$Mg. Similar TS could exist in other highly prolate light nuclei.
They could serve as promising tests for various reactions to probe
a vortical (toroidal) nuclear flow.
\end{abstract}

\maketitle

The toroidal \cite{Dub75,Se81} dipole resonance (TDR) in nuclei
was predicted about 40 years ago and since that time it is a subject
of a permanent high interest, see  \cite{Co00,Vr02,Pa07} for early
and \cite{Kv11,Rep13,Rei14,NePAN16,Rep17} for recent self-consistent
studies. The resonance demonstrates a variety of intriguing properties.
i) Its computed nuclear current \cite{Se81,Bas93,Mis06,Ry02,Rep17} forms a torus-shaped
vortex ring where the current exhibits small-amplitude vortical oscillation around
a closed loop line, see Fig. 1(a,b). It reminds so-called spherical
Hill's vortex  well known in hydrodynamics
\cite{Hill1894,MiTho60}. The difference is that in
Hill's vortex the flow spins around the torus while in  TDR the current
oscillates along the same streamlines.
ii) TDR is the only known example of the {\it intrinsic electric
vortical} flow in nuclei \cite{Se81,Pa07,Rei14}. The toroidal
strength can  be used as measure of the nuclear dipole vorticity \cite{Rei14}.
iii) TDR is the transversal "zero sound" mode with the
features of  an {\it elastic} medium \cite{Bas93,Mis06}.
iv) TDR perhaps is the source of so-called
pygmy dipole resonance \cite{Rep13,NePAN16}.
v) TDR is coupled \cite{Vre_PLB00}
to the irrotational compression dipole resonance (CDR) \cite{Ha77,St82}.
Furthermore, since TDR is located near particle emission thresholds, it can affect
reaction rates important for nucleosynthesis \cite{Rich04,Mar12}.
Toroidal modes and moments are discussed in solid state physics
\cite{Dub90},  areas of metamaterials, plasmonics and nanophotonics
(see \cite{Ivan17} and refs. therein),  neutron stars \cite{Ba02},
physics of anapole and dark matter \cite{Chiu13}.
Altogether, toroidal modes are of relevance to various areas of physics.

Though TDR and CDR are {\it second-order} dipole modes  (i.e.
related to $r^3 Y_{1m}$ field instead of $r Y_{1m}$ \cite{Kv11}),
they dominate in the isoscalar (T=0) dipole channel of nuclear
excitations and are supposed to constitute the low- and
high-energy parts of the isoscalar giant dipole resonance (IS-GDR)
\cite{Harakeh_book_01,Co00,Vr02,Pa07} observed in $(\alpha,\alpha')$
reaction \cite{YoungexpSn-Pb,YoungexpZr,Uch04}.
Following our recent calculations \cite{NePAN16,Rep17},
the usual interpretation of the toroidal part of IS-GDR experimental data
can be disputed. So, identification of the TDR is still an open
problem. This substantially complicates exploration of toroidal excitations.
\begin{figure}
\includegraphics[width=8.5cm]{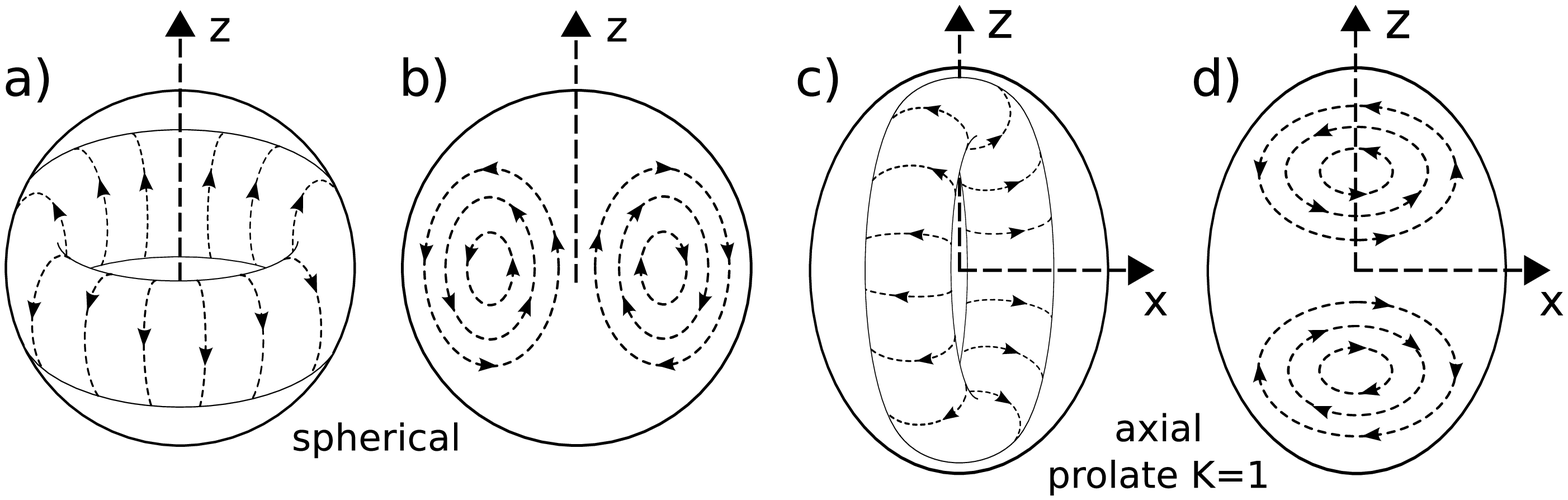}\label{fig1}
\caption{Schematical toroidal dipole flow in spherical  nuclei
 (a-b) and axial nuclei in states with K=1 (c-d).
a) Current lines (arrows) lie on the torus surface
with the axial z-axis determined by an external field.
b) Toroidal flow in a plane embracing z-axis.
c) Toroidal flow for K=1 states (for K=0 states the flow is like in
the plot a)).
d) Toroidal flow for K=1 states in z-x plane.}
\end{figure}

In this connection, we propose a new route to study the toroidal mode:
to switch the experimental and theoretical effort from TDR
(embracing many states and masked by other multipole modes
\cite{YoungexpSn-Pb,YoungexpZr,Uch04})
to {\it individual} well-separated low-energy toroidal states (TS). As shown below,
such TS can exist in low-energy spectra of light nuclei with
a strong axial prolate deformation. What is of a crucial importance,
these states can be easily discriminated and identified.
Using individual TS, we could get new possibilities
for testing various reactions as probes of a nuclear vortical flow.
Besides we could get the answer to the important
question of a general interest: how does Hill's vortex look in the
limit of strongly deformed axial confinement. As shown below, the
torus vortex ring (Fig. 1c-d) is converted to the vortex-antivortex pair.

This idea can be realized using specific deformation features of the isoscalar toroidal mode.
As shown in \cite{NePAN16,Rep17},  the prolate axial deformation splits the mode
into K=0 and K=1 branches and downshifts K=1 strength to form a strong
low-energy toroidal K=1 peak.
The effect can be especially strong in $^{24}$Mg with its
extremely large axial prolate deformation $\beta$=0.605$\pm$0.008 \cite{Raman01} and
a sparse spectrum below the particle thresholds. Just these two factors
offer a chance to find an individual low-energy E1(K=1) TS.

Our calculations are performed within the self-consistent quasiparticle
random-phase-approximation (QRPA) method based on the Skyrme functional
\cite{Ben03}. The QRPA code for axial nuclei \cite{Repcode} employs
2D mesh in cylindrical coordinates.  The calculation box extends over
three times the nuclear radius. The mesh size is 0.4 fm. The single-particle
spectrum embraces all levels from the bottom of the potential well
up to +55 MeV. The volume monopole pairing is treated at the BCS level
\cite{Rep17}. We implement a representative set of
Skyrme parameterizations (SLy6 \cite{SLy6}, SVbas \cite{SV} and SkM* \cite{SkMs})
with various nuclear matter features. SLy6 is used as a main tool
since it gives a good description of the isovector
giant dipole resonance (IV-GDR) in medium-heavy \cite{Kl08}
and light \cite{Kv12_IJMFE} nuclei, including $^{24}$Mg \cite{Supp}.

Various models \cite{Be08,Rod10,Yao11,Hino11,Kimura12} predict
a weak triaxial softness in the ground state of $^{24}$Mg and more
triaxiality in its positive-parity excitations. Concerning the lowest
dipole states \cite{Kimura12}, the triaxiality was found essential for
the K=0 but negligible  for K=1. Here we mainly address the lowest
dipole K=1 state and  effects caused by very large {\it axial} prolate
deformation. So, to simplify the problem, we base our QRPA
analysis on the axial shape of $^{24}$Mg.
The equilibrium axial quadrupole deformation is obtained by
minimization of the total nuclear energy in the ground state. This gives
deformation parameters $\beta$=0.536 (SLy6), 0.525 (SVbas), and 0.493 (SkM*)
in a good agreement (best for SLy6) with the value
$\beta$=0.605$\pm$0.008 obtained from $B(E2)\uparrow$ measurements \cite{Raman01}.

Since the toroidal and compression modes are coupled \cite{Vre_PLB00},
we consider both of them.
Their responses are treated in terms of the reduced transition probabilities
$
B_{\nu}(E1K, \alpha)=(2-\delta_{K,0})|\:\langle\nu|\:\hat{M}_{\alpha}(\;E1K)\:|0\rangle \:|^2
$
where $|\nu\rangle$ is the  wave function of the $\nu$-th QRPA dipole state.
The  toroidal ($\alpha$ = tor) and compression ($\alpha$=com) transition
operators  have the form \cite{Kv11,Rep13}
\begin{equation}\label{TM_curl}
\hat{M}_{\text{tor}}(E1K) = \frac{-1}{10 \sqrt{2}c} \int d^3r r [r^2+d^s+d^a_K]
{\bf Y}_{11K} \cdot ( \bf{\nabla}\!\times\!\hat{\bf j}),
\end{equation}
\begin{equation}\label{CM_div}
\hat{M}_{\text{com}}(E1K) =  \frac{-i}{10c}\int d^3r r[r^2+d^s-2d^a_K]
Y_{1K} (\bf{\nabla} \cdot \hat{\bf j}),
\end{equation}
where $\hat{\bf j}(\bf r)$ is operator of the isoscalar nuclear current
involving convection and magnetization parts (with effective charges
$e_{\rm n,p}^{\rm eff}=0.5$, g-factors $g_{\rm n,p}^s$=0.88, and quenching
 q=0.7 \cite{Kv11,Harakeh_book_01});
${\bf Y}_{11K}(\hat{\bf r})$ and $Y_{1K}(\hat{\bf r})$ are vector
and ordinary spherical harmonics;
$d^s= - 5/3 \langle r^2\rangle_0$ is the center-of-mass corrections (c.m.c.)
in spherical nuclei \cite{Kv11};
$d^a_K = \sqrt{4\pi/45}\langle r^2 Y_{20} \rangle_0 (3\delta_{K,0}-1)$
is the additional c.m.c. arising in axial deformed nuclei within the prescription
\cite{Kv11}. The average values are $\langle f \rangle_0 = \int\:d^3r f \:\rho_0 /A$
where  $\rho_0$ is the g.s. density.
As was checked, the c.m.c. remove spurious admixtures with very high accuracy.
\begin{figure}
\includegraphics[width=7cm]{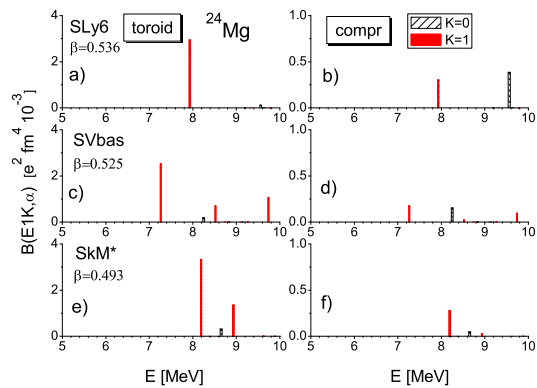}\label{fig:2}
\caption{Toroidal (left) and compression (right) $B(E1,\alpha)$-values in $^{24}$Mg,
calculated with SLy6 (top), SVbas (middle) and SkM* (bottom) forces.
Responses for K=0 (K=1) are depicted by sparse black
( filled red) bars. Deformation parameters are shown for each force.
}
\end{figure}

The toroidal operator with the curl $\bf{\nabla}\!\times\!\hat{\bf j}$
is vortical while the compression operator with the divergence
$\bf{\nabla}\!\cdot\!\hat{\bf j}$ is irrotational. Using the continuity
equation, the current-dependent operator (\ref{CM_div}) can be transformed
\cite{Kv11} to the familiar density-dependent form \cite{Harakeh_book_01}
$
\hat M'_{\text{com}}(E1K) = 1/10 \int d^3r r \hat{\rho} [r^2+d^s-2d^a_K ] Y_{1K}
$
where $\hat{\rho}(\bf r)$ is the density operator. This form 
is often used as a probe field in the analysis of
($\alpha, \alpha'$)-reaction data for IS-GDR \cite{Harakeh_book_01}.
The calculated TDR and CDR strengths constituting IS-GDR
in $^{24}$Mg are shown in \cite{Supp}.

The QRPA calculations use a large two-quasiparticle (2qp) basis with the energies
until $\sim$ 100 MeV. For SLy6, the basis includes $\approx$ 1900 ($K=0$) and
$\approx$ 3600 (K=1) states. For the photoabsorption, the Thomas-Reiche-Kuhn sum rule
\cite{Ring_book,Ne08} is exhausted by 100\% (SLy6, SVbas) and 97\% (SkM*).
The isoscalar dipole energy-weighted sum rule \cite{Harakeh_book_01} is exhausted
by $\approx$ 97 $\%$.
\begin{figure}
\includegraphics[width=7cm]{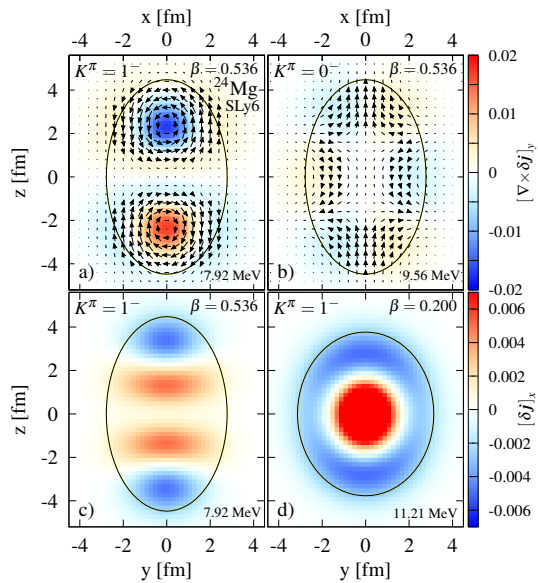}\label{fig3}
\caption{(a-b): QRPA (SLy6) isoscalar dipole CTD $\delta{\bf j}_{zx}$ in z-x (y=0) plane
for the states K=1 at E=7.92 MeV (left) and K=0 at E=9.56 MeV (right).
Magnitude of $\delta {\bf j}_{zx}$ is determined by arrow lengths in arbitrary units.
The vorticity $({\bf \nabla} \times \delta\bf{j})_y$ is shown by colors as indicated.
(c-d) The CTD  $\delta {\bf j}_x$ in z-y (x=0) plane for K=1 states computed at
deformations $\beta$=0.536 (left, E=7.92 MeV) and
$\beta$=0.2 (right, E=11.21 MeV). Unlike plots (a-b), the CTD  are shown
by colors. In all the plots, the nuclear surface is indicated by the solid line.}
\end{figure}

In Fig. 2, the calculated  toroidal ($\alpha$ = tor) and compression ($\alpha$ = com) values
$B(E1K,\alpha)$ are shown for  $I^{\pi}=1^-$ QRPA states  at the energy 5$ < E < $10 MeV
(at $E <$5 MeV our calculations do not give $1^-$ states).
The  transition operators (\ref{TM_curl}) and (\ref{CM_div})  are used.
Plot a) for SLy6 shows that the toroidal strength has an
impressive K=1 peak at 7.92 MeV. What is important, this state is the
{\it lowest} $1^-$ excitation.
In the plot b), the irrotational compression strength is
given by 7.92-MeV K=1 state and subsequent 9.56-MeV K=0 state (lowest among
K=0 excitations). The exposure of the toroidal 7.92-MeV K=1 state in the compression response
means that this state has a minor irrotational fraction.
The middle and bottom plots of Fig. 2 show the responses calculated with
SVbas and SkM*. The results noticeably deviate in details from SLy6 picture.
However they also give the lowest toroidal K=1 state (at 7.26 MeV  for SVbas and
8.19 MeV for SkM*) accompanied by the subsequent compression K=0 state.
So the main result, individual toroidal K=1 state as the
lowest $1^-$ excitation, is robust. The similar result was recently
obtained within the Antisymmetrized Molecular Dynamics (AMD) approach
in strongly deformed $^{10}$Be \cite{Kanada17}.

The toroidal character of 7.92-MeV K=1 state is additionally justified
in Fig. 3a)  where its isoscalar current transition density (CTD)
$\delta{\bf j}_{zx}=\langle \nu | (\hat{\bf j}_z + \hat{\bf j}_x)|0\rangle$
at the plane z-x (y=0) is shown. For simplicity,
only convection part of the nuclear current is taken into account.
Note that CTD by definition is fully determined by the structure of the QRPA $\nu$-state
and does not depend on the transition operators (\ref{TM_curl})-(\ref{CM_div}).
Plot a) shows that the flow in K=1 state is indeed toroidal. Like in the
schematic picture of Fig. 1d), its streamlines spin with opposite circulations
around two parallel axes (perpendicular to z-x plane) in the top and bottom
of the prolate system. At the same time,  Fig. 3a) shows some important peculiarities. Unlike
the toroidal pattern  in spherical nuclei \cite{Rep13,NePAN16,Ry02} and prolate $^{154}$Sm \cite{Rep17}
where the flow fills in most of the nuclear volume, here the flow is concentrated in the top and
bottom of the nucleus and is almost absent in the equatorial region (-0.8 fm $<$ z $<$ 0.8 fm).
So, in highly deformed $^{24}$Mg, the torus-shaped vortex ring converts into well separated vortex
and antivortex. Both them are characterized by a high vorticity $({\bf \nabla} \times  \delta{\bf j})_y$.
Similar current fields take place for 7.26-MeV  and 8.19-MeV K=1 states in SVbas and SkM* (not shown).

The character of the flow is even more clarified in the plot c) were CTD
$\delta{\bf j}_x$ in z-y (x=0) plane embracing the vortex/antivortex axes
is shown. We see that CTD indeed has opposite directions around
z$\approx$ 3.5 fm (blue) and 1.5 fm (red) in the top. The same is around -3.5 fm
(blue) and -1.5 fm (red) in the bottom. The flows are maximal in these regions
and vanish in the equatorial area, which corresponds to well separated vortex and
antivortex.  What is important, the flow in the plot c) does not form two elliptic
areas (red surrounded by blue) like in the plot d) where  CTD
$\delta{\bf j}_x$ is shown for  the toroial 11.21-MeV K=1 state  obtained at the small constrained
deformation $\beta$=0.2. This proves that the current field in 7.92-MeV K=1 state
of $^{24}$Mg  indeed acquires the vortex-antivortex configuration and
that the high prolate deformation in  $^{24}$Mg is crucial for this transformation.
By our knowledge, this result  is the first demonstration of the
transformation of the spherical Hill's vortex \cite{Hill1894,MiTho60} in a highly
deformed confinement.
\begin{figure}
\includegraphics[width=8.5cm]{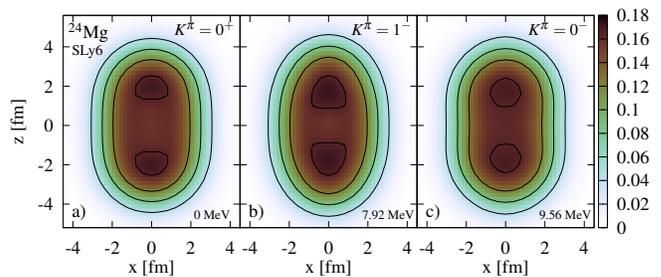}\label{fig4}
\caption{QRPA (SLy6) densities (in fm$^{-3}$) of the ground state (a), 7.92-MeV K=1 state
(b) and 9.56-MeV K=0 state (c).}
\end{figure}

The toroidal 7.92-MeV K=1 state is mainly formed by two 2qp components,
$pp [211\uparrow-330\uparrow]$ and $nn [211\uparrow-330\uparrow]$ (in Nilsson
asymptotic quantum numbers), exhausting
54$\%$ and 39$\%$ of the norm, respectively. The comparison of Fig. 4a with the
currents for these components  (not shown) reveals that the toroidal flow is
predominantly of 2qp origin. This agrees with the previous finding for $^{208}$Pb
\cite{Dresden}.

For the comparison, in Fig. 3b)  the CTD for K=0 state at 9.56 MeV is given.
Its flow is basically irrotational. It is concentrated in four regions near the nuclear
surface and vanishes at the nuclear center. The "static" intervals between
these regions can be treated as compression (z$\sim$ -2 fm) and decompressions
(z$\sim$ 2 fm) areas. This suggests that 9.56-MeV K=0 state is a compression mode.

The nucleus $^{24}$Mg is known to demonstrate cluster properties
\cite{But96,clust_rev,Chi16}. Energies of 7.92-MeV K=1 and  9.56-MeV K=0 states  are
close to the $\alpha$-particle threshold $S_{\alpha}$= 9.3 MeV \cite{bnl} and so
these states can be related to cluster degrees of freedom. A significant
separation of the vortex and antivortex in Fig. 3a) (in contrast to the toroidal flow
covering all the nucleus interior in $^{154}$Sm \cite{Rep17}) can also signal on the cluster
structure. To check this, we show in  Fig. 4 density distributions
$\langle \nu|{\hat\rho}|\nu\rangle$ in z-x (y=0) plane
for the  ground state and QRPA states 7.92-MeV K=1 and  9.56-MeV K=0.
In all these states, there are cumulations of density at the top and bottom
of the nucleus, which can be interpreted as precursors of clustering, e.g. of $\alpha$ formation.
The form of the cumulation regions and separation between them are somewhat different.
Anyway we see that $^{24}$Mg demonstrates a similar clustering in the ground state and
dipole excitations around $S_{\alpha}$. Perhaps this is a manifestation of mean-field/cluster
duality pertinent to light nuclei \cite{Chi16}.  The clustering
can favor a significant spatial vortex/antivortex separation in the toroidal K=1 state
and signify a possible $\alpha$-vibrational contribution to the irrotational K=0 state.
These aspects deserve a further investigation with more decisive observables \cite{Re11}.

\begin{figure}
\includegraphics[width=8cm]{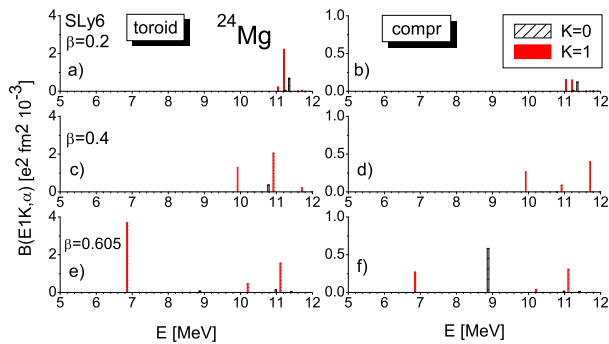}\label{fig5}
\caption{Toroidal (left) and compression (right) $B(E1,\alpha)$-values in $^{24}$Mg,
calculated with SLy6 for constrained deformations $\beta_2$=0.2 (upper), 0.4 (middle)
and 0.605 (bottom). For each case, the $K=0$ (sparse black bars),
and $K=1$ (filled red bars) states are depicted.}
\end{figure}

A crucial role of the large axial prolate deformation is illustrated in Fig. 5.
We see that, for the small deformation $\beta$=0.2,  K=1 toroidal state lies at 11.2 MeV
and is not the lowest one. At the larger deformation $\beta$=0.4, the toroidal strength is
fragmented between two K=1 peaks where the lowest come down already to 9.9 MeV. Finally, at
 the experimental value $\beta$=0.605 \cite{Raman01}, the toroidal K=1 strength is mainly
concentrated in the lowest peak at 6.9 MeV, a feature which is well developed
already at $\beta$=0.536 in Fig. 2a). Besides, at $\beta$=0.605, we get a concentration
of the compression strength at 8.9 MeV. Altogether we see that the toroidal K=1
peak becomes the lowest and well separated state only at a high prolate deformation.

The toroidal 7.92-MeV K=1 state exhibits  some collectivity.
Its collective downshift (the difference between the QRPA energy and energy
of the largest 2qp component) is 0.85 MeV, i.e. rather large.  Besides, the state
shows a strong E3 transition with
$B(E3, 0^+0_{\rm g.s.} \to 3^-1)$ = 402 $\rm{e^2 fm^6}$ (11.7 W.u.), which can
be caused by the deformation-induced mixture of dipole and octupole modes
and significant octupole correlations \cite{Noto81}.
Though the state is mainly isoscalar, it also has a minor isovector
fraction resulting in the isovector E1 decay to the ground state (g.s.) band:
$B(E1, 1^-1 \to 0^+0_{\rm g.s.})$ = 2.52 $10^{-4} \rm{e^2 fm^2}$.
The experimental spectrum below 10 MeV in $^{24}$Mg has three $I^{\pi}=1^-$
states: at 7.555, 8.437 and 9.566 MeV \cite{bnl}. Their K-assignment is rather
ambiguous.  Following Alaga rules \cite{Ala55},
the toroidal 7.92-MeV K=1 state might correspond to  7.555-MeV or
8.437-MeV levels, see more details in \cite{Supp}.

In conclusion, our Skyrme QRPA calculations show that, in highly deformed axial  $^{24}$Mg,
the {\it lowest} dipole K=1 state is a vortical toroidal excitation.
Its computed energy (7.92 MeV  for the Skyrme force SLy6) is close to the energies
7.555 and 8.437 MeV of two lowest $1^-$  levels in the experimental spectrum.
This state is well separated from the surrounding spectrum and so represents the
example of an {\it individual} toroidal state TS). Following our analysis, TS becomes
lowest due to the exceptionally high axial prolate deformation ($\beta_{\rm exp}$=0.605) in $^{24}$Mg.
Perhaps, individual low-energy TS also exist in other strongly
deformed light nuclei, e.g. in $^{10}$Be \cite{Kanada17}, $^{20,22,24}$Ne and $^{32,34}$Mg.
N=Z nuclei, like $^{24}$Mg, are most promising since they do
not have pygmy dipole admixtures.

The lowest energy of the K=1 TS in $^{24}$Mg greatly simplifies its identification.
TS should have peculiarities in $(e,e')$ back scattering \cite{Rich04}.
TS could serve as excellent test cases to probe various reactions
for vortical nuclear excitations.

The toroidal nature of 7.92-MeV K=1 state is proved by a large
toroidal strength and clear toroidal distribution of the nuclear
current. Actually we get so-called Hill's vortex well
known in hydrodynamics of turbulent fluids \cite{Hill1894,MiTho60}.
Here we show for the first time, that, in a highly deformed axial confinement,
Hill's vortex is transformed from toroidal to vortex-antivortex configuration.

   The calculated nuclear density shows two clear well-separated cluster  regions
matching Hill's vortex and antivortex areas. The clustering takes place
in the ground state and dipole toroidal K=1 and compressions K=0 states.
Perhaps here we have an interesting interplay of clustering and vortical mean-field
dynamics.

Main results persist for different Skyrme forces (SLy6, SV-bas, SkM*).
Since the results are mainly caused  by a huge axial prolate deformation,
they can hardly be essentially changed by a possible triaxiality  and coupling with
complex configurations. Anyway we plan to scrutinize these effects
in the subsequent studies.

V.O.N. thanks Profs. M.N. Harakeh and A. Diaz-Torres for valuable discussions.
The work was partly supported by the
Heisenberg - Landau (Germany - BLTP JINR), and Votruba - Blokhintsev (Czech Republic
- BLTP JINR) grants. A.R. is grateful for support from Slovak Research and Development
Agency (Contract No. APVV-15-0225) and Slovak grant agency VEGA (Contract No. 2/0129/17).
J.K. appreciates the support of the research plan MSM 0021620859 (Ministry of Education of the
Czech Republic) and the Czech Science Foundation project P203-13-07117S.%

\end{document}